\documentclass[prc,superscriptaddress,unsortedaddress,twocolumn,showpacs,preprintnumbers,amsmath,amssymb]{revtex4}
\usepackage[dvips]{graphicx}
\usepackage{amsmath}
\usepackage{dcolumn}

\begin{document}
\title{ Determination of $S_{17}$ from $^7$Be($d,n$)$^8$B reaction:\\
CDCC analyses based on three-body model
}

\author{Kazuyuki Ogata}
\email[Electronic address: ]{kazu2scp@mbox.nc.kyushu-u.ac.jp}
\affiliation{Department of Physics, Kyushu University,
Fukuoka 812-8581, Japan}
\author{Masanobu Yahiro}
\affiliation{Department of Physics and Earth Sciences,
University of the Ryukyus, Nishihara-cho, Okinawa 903-0213, Japan}
\author{Yasunori Iseri}
\affiliation{Department of Physics, Chiba-Keizai College,
Todoroki-cho 4-3-30, Inage, Chiba 263-0021, Japan}
\author{Masayasu Kamimura}
\affiliation{Department of Physics, Kyushu University,
Fukuoka 812-8581, Japan}

\date{\today}

\begin{abstract}
    The astrophysical factor $S_{17}$ for $^7$Be($p,\gamma$)$^8$B reaction
is reliably extracted from
the transfer reaction $^7$Be($d,n$)$^8$B at $E=7.5$ MeV
with the asymptotic normalization coefficient method.
The transfer reaction is accurately analyzed with CDCC
based on the three-body model.
This analysis is free from uncertainties of the optical potentials
having been crucial in the previous DWBA analyses.
\end{abstract}


\maketitle

The solar neutrino problem is a central subject in
the neutrino physics~\cite{Bahcall}.
The major source of the high-energy neutrinos observed by solar neutrino
detectors is $^8$B produced by the $^7$Be($p,\gamma$)$^8$B reaction.
The astrophysical factor $S_{17}$ for the reaction, however, is one of
the most poorly determined reaction rates in the standard solar model;
the latest recommendation for the factor $S_{17}(0)$ at zero energy,
based on recent direct measurements,
is $19^{+4}_{-2}$ eV b~\cite{Adelberger}, i.e.,
10--20 \% error exists.
This is far from our goal of determining $S_{17}(0)$ within 5 \% error
required in order to determine the neutrino oscillation parameters:
the mass difference between $\nu_{e}$ and $\nu_{\mu}$ and their mixing
parameter. The main difficulty in the direct measurement comes from
ambiguities of determining the effective target thickness of the
radioactive $^7$Be beam.
Thus, indirect measurements of $S_{17}(0)$ are expected to be essential
for determining $S_{17}(0)$ accurately.

The transfer reaction $^7$Be($d,n$)$^8$B at low energies is an example of
such indirect measurements;
once the asymptotic normalization coefficient
(ANC) of the overlap function between the $^7$Be and $^8$B ground states
is determined from the data of the transfer reaction,
$S_{17}(0)$ can be accurately derived from the ANC,
as long as the reaction is peripheral~\cite{Xu}.
The reaction has been measured at $E=7.5$ MeV and the ANC is
extracted with the distorted wave Born approximation (DWBA)~\cite{Liu}.
The  $S_{17}(0)$ obtained with the ANC is $27.4 \pm 4.4$ eV b,
leading to inconsistency with the recommended value.
Although the reaction is found with DWBA to be indeed
peripheral~\cite{Gagliardi,Fernandes},
distorting potentials used in DWBA are quite
ambiguous~\cite{Gagliardi,Fernandes}, which prevent one from extracting
$S_{17}(0)$ accurately.
In particular, uncertainties of the $d$-$^7$Be optical potential
bring about large errors for $S_{17}$, typically 30 \% in magnitude.
The origin of the large ambiguity of distorting potentials is
that these are derived from proton and deuteron optical potentials
for different targets and/or energies.

In the present paper, we analyze $^7$Be($d,n$)$^8$B at 7.5 MeV
with the three-body model, $p+n+^7$Be, assuming $^7$Be to be an inert core.
This treatment of the system is justified by Refs.~\cite{Nunes1,Nunes2},
where core excitations of $^7$Be are shown to be negligible.
An advantage of this analysis is that
we do not need the ambiguous $d$+$^7$Be optical potential in the entrance
channel and the ambiguous $n$+$^8$B one in the exit channel as shown
below.

The three-body dynamics in the entrance channel are explicitly treated
by means of continuum-discretized coupled-channels (CDCC)
method~\cite{CDCC,Piya}, the theoretical foundation of which is given
in Ref.~\cite{CDCC-foundation}.
This theory has been established as a method of solving the three-body
system with good accuracy, and extensively applied for various
reactions~\cite{CDCC,Surrey}.
Previous CDCC calculation showed that explicit treatment of breakup
channels is essential in describing deuteron induced reactions~\cite{CDCC}.
The CDCC thus provides a precise description of the wave function
in the entrance channel, i.e., $d$-$^7$Be system.

 The effective Hamiltonian for the entrance channel
based on the three-body model contains
the optical potential between $N$ ($p$ or $n$) and $^7$Be.
Data of the neutron elastic scattering are available for target
$^7$Li, the mirror nucleus of $^7$Be, at 4 MeV, approximately half the
deuteron energy considered here~\cite{n+A.dat}.
First the $n$-$^7$Li potential is determined accurately from the data.
The potential is then used as an input
in CDCC calculation for deuteron elastic scattering on $^7$Li at 8 MeV,
and the numerical result is compared with the experimental
data~\cite{d+A.dat}.
This is a good test for the neutron optical potential determined above,
which is only an input in CDCC calculation for the entrance channel.
As for the exit channel, the three-body dynamics are
treated with the adiabatic approximation, after testing its accuracy
with CDCC calculation. The ANC is then obtained with reliable distorted
wave functions in both the entrance and exit channels.

    The transition amplitude for the transfer reaction,
based on the three-body model ($p+n+^7$Be), is
\begin{equation}
   T_{\rm fi}= S_{\rm exp}^{1/2} < \Psi_{\rm f}^{(-)} |
   V_{\rm np}
   | \Psi_{\rm i}^{(+)} > .
\label{T}
\end{equation}
The three-body wave function $\Psi_{\rm i}^{(+)}$ is a solution of
the Schr\"{o}dinger equation $( H_{\rm i}-E ) \Psi_{\rm i}^{(+)}=0$
for the three-body Hamiltonian
\begin{equation}
 H_{\rm i} = K_{\rm np} + K_{\rm dBe}   +  V_{\rm np}({\bf r}_{\rm np})
   + U,
\label{H-init}
\end{equation}
with $U=U_{\rm nBe}({\bf r}_{\rm nBe}) + U_{\rm pBe}({\bf r}_{\rm pBe})$.
Here ${\bf r}_{\rm XY}$ is
the coordinate of nucleus X relative to nucleus Y.
The potential $V_{\rm np}$ is the interaction between $n$ and $p$,
$U_{\rm pBe}$ ($U_{\rm nBe}$) is the proton (neutron) optical
potential for the target $^7$Be at half the  deuteron incident energy,
and $K_{\rm np}$ and $K_{\rm dBe}$ show kinetic energy operators for two-body
systems denoted by the subscripts.
The nuclear part of $U_{\rm pBe}$ is assumed to be the same as $U_{\rm nBe}$.
The Coulomb part of $U_{\rm pBe}$ is treated approximately by
replacing the coordinate $r_{\rm pBe}$ by $r_{\rm dBe}$.
Effects of Coulomb breakup of deuteron, not included in the
usual treatment above, are found to be quite small
in the present system.

The wave function $\Psi_{\rm i}^{(+)}$ is obtained with CDCC, that is,
by solving the Schr\"{o}dinger equation in a model space.
In CDCC, deuteron breakup states are classified with
linear and angular momenta, $k$ and $\ell$, and
truncated into $0 \le k \le k_{\rm max}$ and $0 \le \ell \le \ell _{\rm max}$,
respectively.
The $k$ continuum $[0,k_{\rm max}]$ is further
divided into bins with a common width $\Delta$.
The total wave function is expanded, in terms of the deuteron ground state
and the discretized breakup states, into
\begin{equation}
\Psi_{\rm i}^{(+)} =
\sum_{i} \phi_i({\bf r}_{\rm np}) \chi_i({\bf r}_{\rm dBe}) ,
\end{equation}
where $\phi_0$ is the deuteron ground state and $\phi_i$ is the $i$-th
discretized breakup state obtained by averaging
continuous breakup states in the $i$-th bin.
The coefficient $\chi_i$ represents a center-of-mass motion of $n$-$p$ pair
in the $i$-th state.
Inserting this form into the three-body Schr\"{o}dinger equation leads to
a set of coupled differential equations,
\begin{equation}
  ( E_i - K_{\rm dBe} ) \chi_i = \sum_{j} F_{ij} \chi_j
\label{eq:CDCC}
\end{equation}
with $F_{ij}=<\phi_i|U|\phi_j>_{{\bf r}_{\rm np}}$ and $E_i=E-e_i$, where
$e_i$ is an intrinsic energy of the $i$-th $n$-$p$ state.
The coupled equations are soluble, since they have a compact kernel in
its integral equation form. The precise formulation of CDCC is shown
in Ref.~\cite{Piya}.
The present model space is $k_{\rm max}=1.7 $ fm$^{-1}$, $\ell=0, 2$
and $\Delta=1.7/40$ fm$^{-1}$.
The CDCC solution converges at these values,
as the model space is enlarged.

The exit channel is also treated in
the three-body model, that is, the exit channel wave function
$\Psi_{\rm f}^{(-)}$ is determined by
the three-body Hamiltonian
\begin{equation}
 H_{\rm f} = K_{\rm pBe} + K_{\rm nB}
 + V_{\rm pBe}({\bf r}_{\rm pBe}) + U_{\rm nBe}({\bf r}_{\rm nBe}).
\end{equation}
In the three-body model, $^8$B is treated
by the two-body ($p$+$^7$Be) model with
the potential $V_{\rm pBe}$.
The spectroscopic factor $S_{\rm exp}$ in Eq. (\ref{T}) is
introduced by taking account of the incompleteness of the model.
It should be noted that $H_{\rm f}$ does not contain $V_{\rm np}$
since the interaction is already treated as a transition operator
in Eq.~(\ref{T}).
In general, the distorting potential $U_{\rm nBe}$ between an outgoing
neutron and $^7$Be differs from the corresponding one in $H_i$,
since an outgoing neutron has a different velocity
from an incoming deuteron in the ($d,n$) reaction.

In principle, $H_{\rm f}$ allows transitions between the ground and
continuum states of the p+$^7$Be system in the exit channel.
However, effects of the transitions
may be simply estimated with the adiabatic approximation,
since the ground state of $^8$B has a binding energy (0.137 MeV)
considerably smaller than an energy (4.18 MeV) of outgoing neutron.
Following Johnson and Soper~\cite{Johnson-Soper},
we replace the Hamiltonian of the p+$^7$Be system
by the binding energy of $^8$B. Errors of the adiabatic approximation
are estimated
with CDCC in the exit channel scattering, $^8$B($n,n$)$^8$B, at $E=4.18$ MeV.
The breakup effect itself is not so large and
errors of the approximation are less than 3 \% at forward angles.
The Johnson-Soper approximation leads to a simple form
$\Psi_{\rm f}^{(-)}=\chi_{\rm nB}^{(-)} \phi_{\rm pBe}$, where
$\phi_{\rm pBe}$ is the wave function of $^8$B in its ground state and
$\chi_{\rm nB}^{(-)}$ is the wave function of outgoing neutron distorted
by the potential $U_{\rm nBe}(r_{\rm nB}\cdot8/7)$, where
the zero-range approximation is made to the transition amplitude (\ref{T}).
We discuss the use of this approximation below.
It should be noted that in DWBA
$\chi_{\rm nB}^{(-)}$ is determined by the elastic scattering of neutron
from $^8$B, of which no measurement has been done so far.
On the other hand, the three-body model approach can avoid this difficulty,
as in the entrance channel.

The transfer reaction $^7$Be($d,n$)$^8$B is calculated with
the zero-range approximation with its finite-range correction~\cite{fr-corr}.
The integration over ${\bf r}_{\rm pBe}$ in $T$ is made up to a large
value 40 fm,
since the transferred proton is very weakly bound in $^8$B.
The finite-range correction for the transition amplitude (\ref{T})
including deuteron ground and breakup channels is straightforward;
the resultant correction for the $i$-th channel keeps
the standard form by regarding
$F_{ii}$ and $e_i$ as the potential and the intrinsic energy
of the entrance channel.
This prescription is tested by doing finite-range DWBA calculation for
the deuteron ground channel which is a main component
of the transition amplitude (\ref{T}).
The result of the prescription above agrees with that of the
full finite-range calculation within 2 \% error at forword angles.

The wave function $\phi_{\rm pBe}$ is calculated with four
types of $V_{\rm pBe}$~\cite{Kim,Tombrello,Robertson,Esbensen}.
We determine the spectroscopic factor $S_{\rm exp}$ comparing the calculated
$^7$Be($d,n$)$^8$B cross section with the experimental one, for every
type of the four potentials.
The astrophysical factor $S_{17}(0)$ at zero energy is then obtained
from the $S_{\rm exp}$ with
\begin{equation}
 S_{17}(0)=\frac{S_{\rm exp} b^2}{0.026}
\label{S17}
\end{equation}
in the ANC method~\cite{Xu}, where $b$ is defined
with the Whittaker function $W$ as $\phi_{\rm pBe}(r_{\rm pBe})
=b W(r_{\rm pBe})$ at $r_{\rm pBe}$ larger than the range of
the nuclear force $V_{\rm pBe}$ between $p$ and $^7$Be.
It should be noted that if the reaction is peripheral,
ANC should be stable against the change of
the $^8$B internal wave functions, i.e., that of the parameter set of
single-particle potentials.
Thus, one can estimate the error of the ANC calculation
from the deviation of ANC (or $S_{17}$) with
four different models of $^8$B above.

\begin{figure}[bht]
\begin{center}
\includegraphics[width=70mm,keepaspectratio]{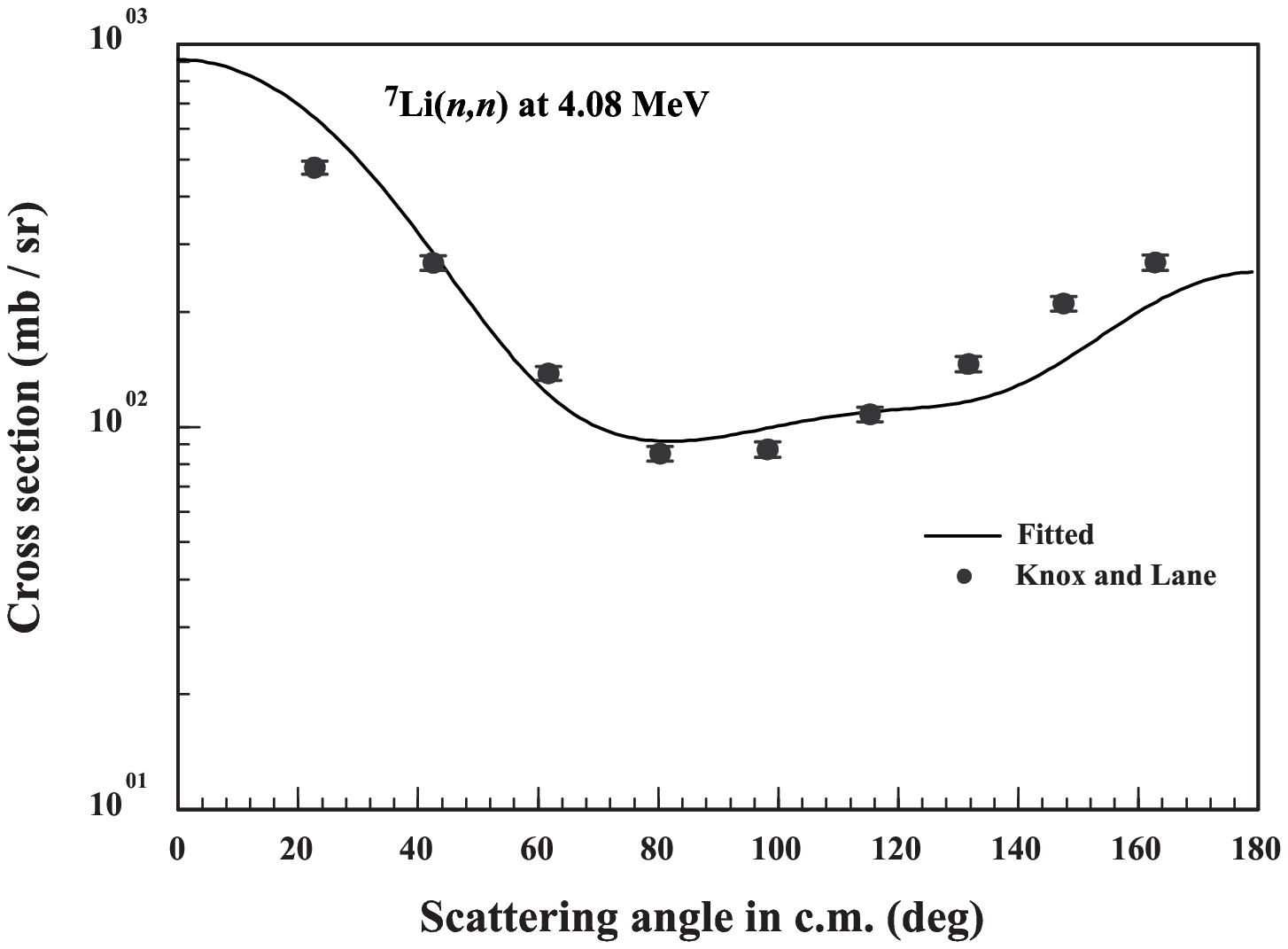}
    \caption{Results of the optical potential search for
             neutron elastic scattering at 4.08 MeV from $^7$Li.
             Experimental data are taken from Ref.~\protect\cite{n+A.dat}.}
\includegraphics[width=70mm,keepaspectratio]{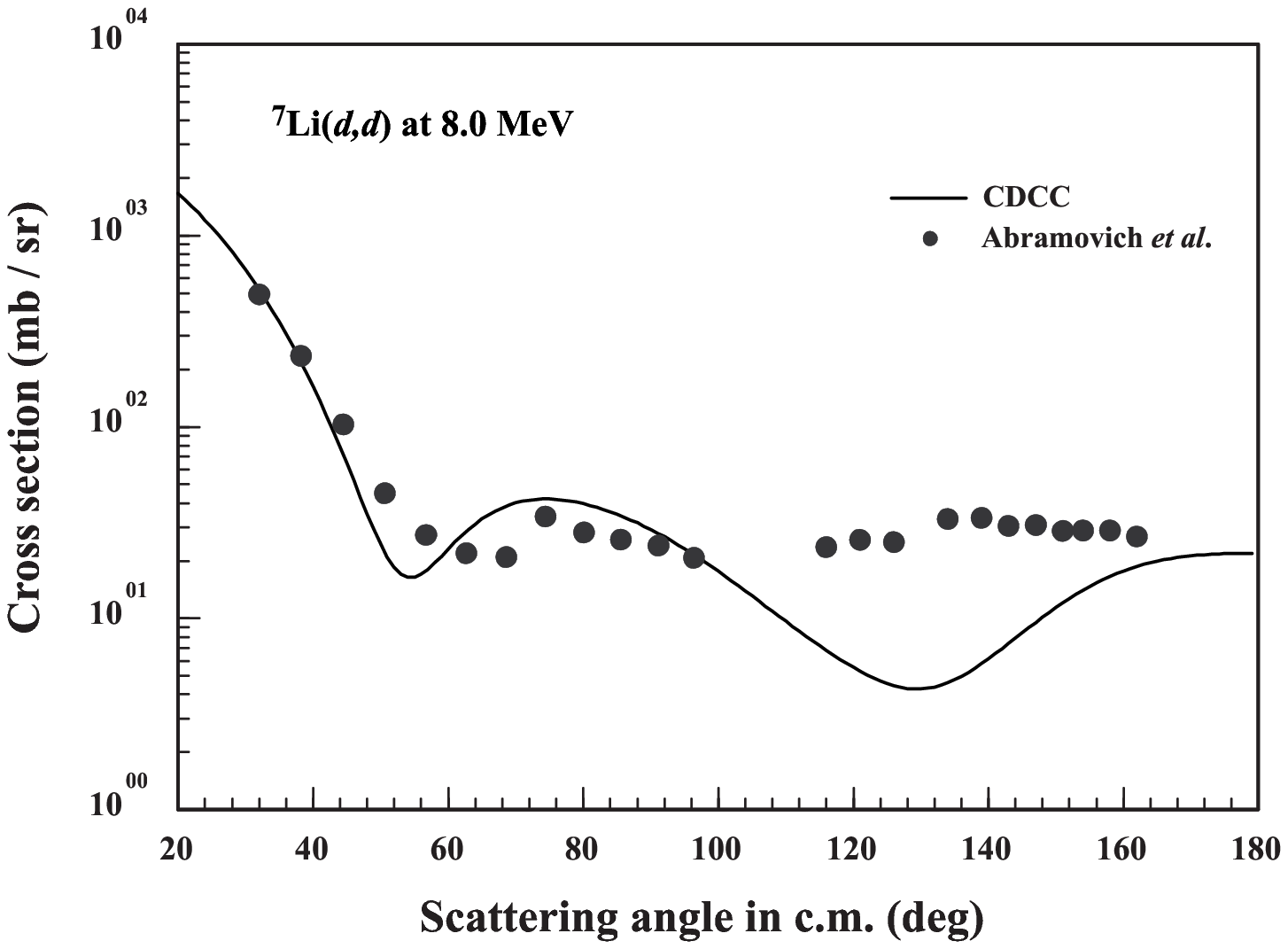}
    \caption{Comparison between the CDCC calculation
             and the
             experimental data~\protect\cite{d+A.dat} for
             $^7$Li($d,d$)$^7$Li at 8.0 MeV.}
\end{center}
\end{figure}
Figure 1 shows the result of the optical potential search for
neutron scattering at 4.08 MeV from $^7$Li,
the mirror nucleus of $^7$Be.
The resultant potential shows a good agreement with data~\cite{n+A.dat}.
The optical potential is then applied for
deuteron scattering at 8.0 MeV from $^7$Li.
The CDCC calculation with the potential again
gives a good agreement with data~\cite{d+A.dat}
at angles $ \theta < 60^\circ $,
as shown in Fig.~2. The potential is shown to be reliable
especially at the forward angles; we use it in $H_i$ and
obtain the proper wave function for the entrance channel.
As for the exit channel, on the other hand,
we need the optical potential of
the $n$+$^7$Li scattering at $E=4.18$ MeV to determine $H_f$.
The data is available at $E=4.26$ MeV that is closest to
the proper energy. Figure 3 shows the result of the optical
potential search for the scattering. The resulting
potential well reproduces the data~\cite{n+A.dat2} at forward angles.
\begin{figure}[ht]
\begin{center}
\includegraphics[width=70mm,keepaspectratio]{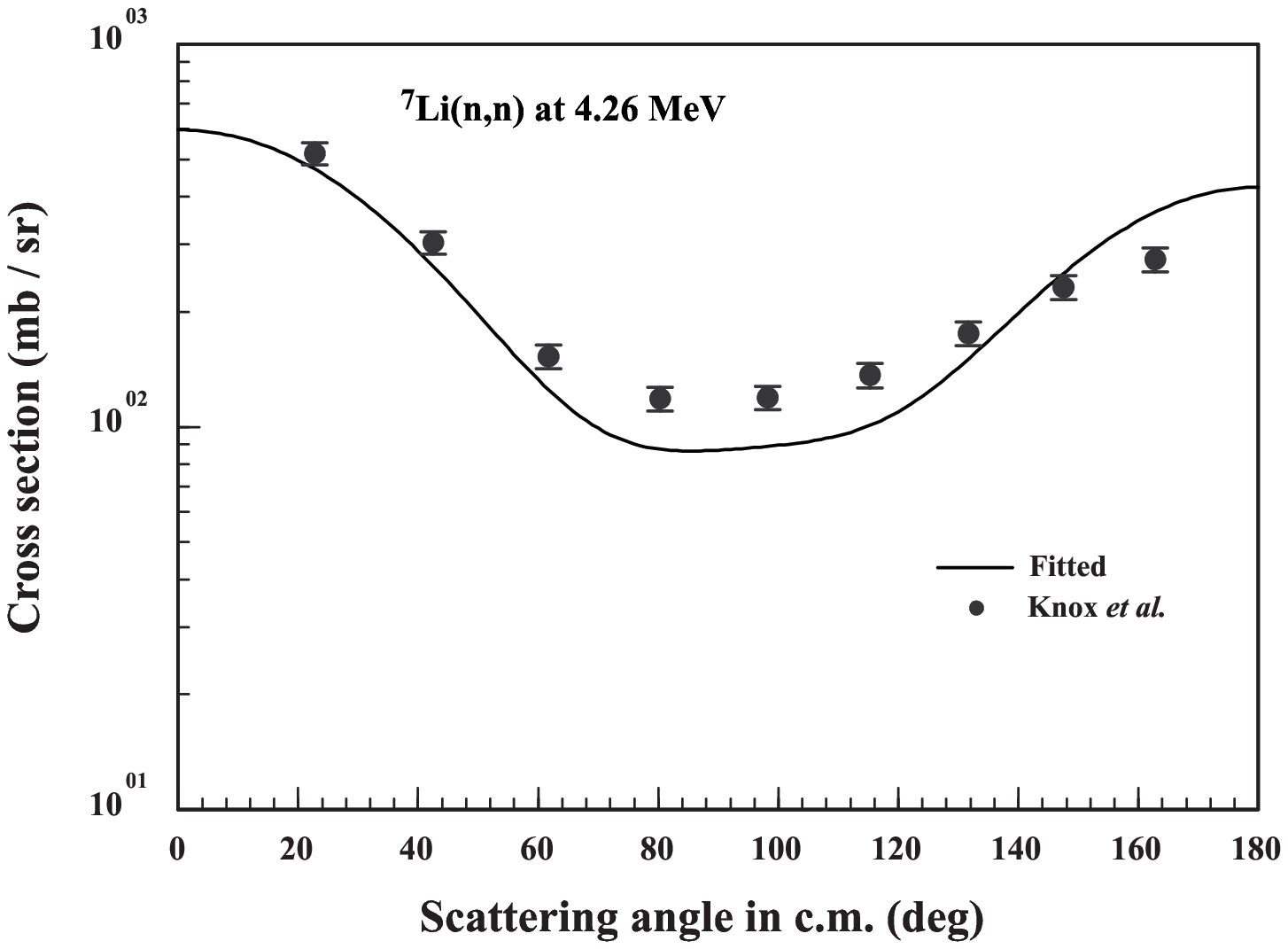}
    \caption{Results of the optical potential search for
             neutron elastic scattering at 4.26 MeV from $^7$Li.
             Experimental data are taken from Ref.~\protect\cite{n+A.dat2}.}
\includegraphics[width=70mm,keepaspectratio]{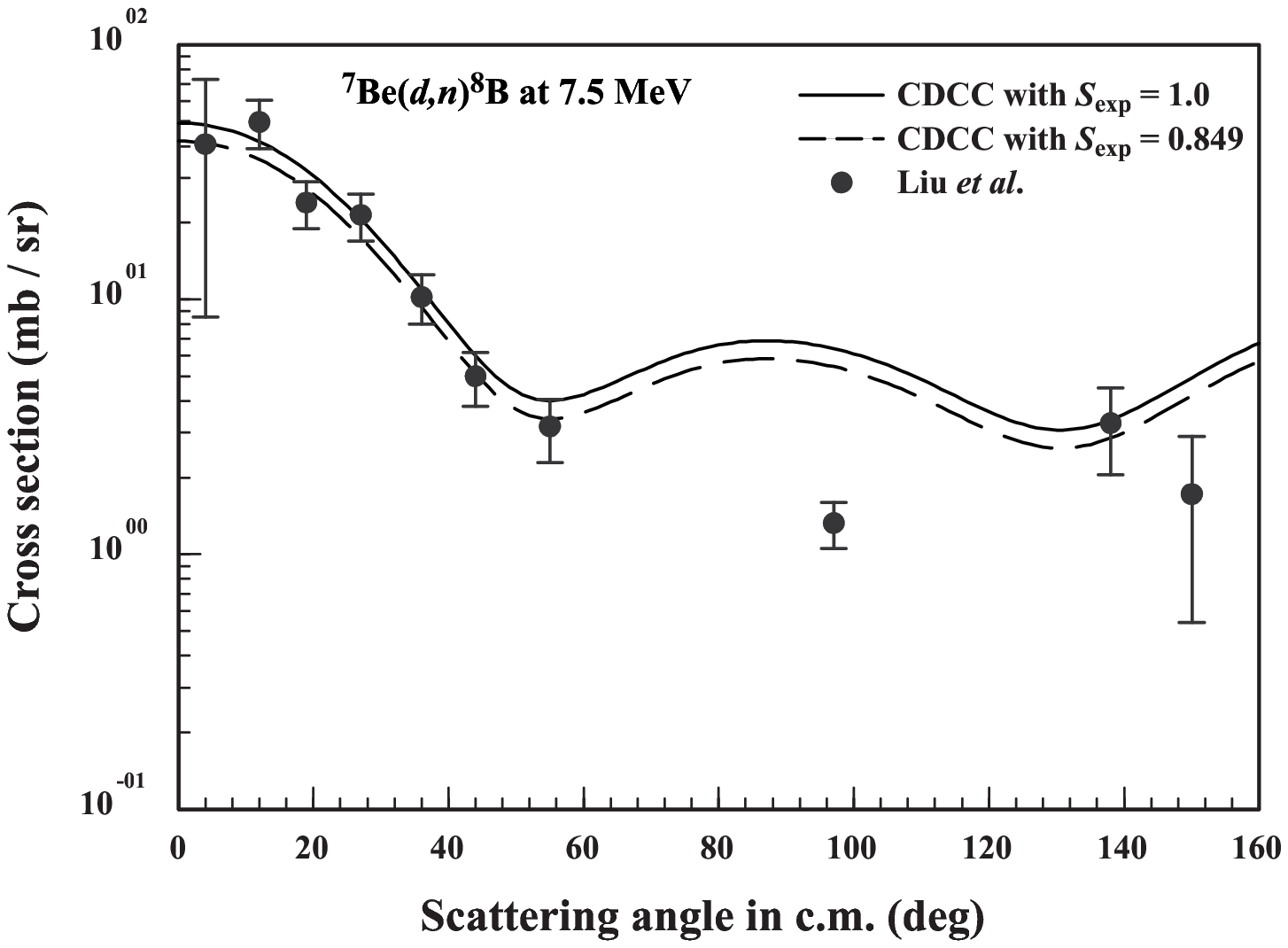}
    \caption{The calculated cross sections for $^7$Be($d,n$)$^8$B
    at 7.5 MeV with $S_{\rm exp}=1.0$ (solid line) and 0.849 (dashed line),
    compared with the experimental data~\protect\cite{Liu}.}
\end{center}
\end{figure}

\begin{table}[ht]
\caption{Parameters for the optical potentials between neutron and $^7$Li
at $E_n$= 4.08 (a) and 4.26 MeV (c) corresponding to the initial and
final channels for $^7$Be($d,n$)$^8$B at $E_d$= 7.5 MeV, respectively.
The single particle potential between $p$ and $^7$Be in $^8$B of
Kim {\it et al}.~\protect\cite{Kim} (b) is also shown.
}
\begin{ruledtabular}
\begin{tabular}{c|ccccccccc}
    &  $V_0$  &  $r_0$  &  $a_0$  &  $W_d$  &  $r_i$
&  $a_i$ & $V_{so}$ & $r_{so}$ & $a_{so}$ \\ \hline
a & 46.57  & 2.07  & 0.49  & 0.82 & 1.87   & 0.22
& 5.50  & 1.15  & 0.50\\
b & 32.12 & 1.54  & 0.52  & ---  & ---    & ---
& 8.24  & 1.54  & 0.52\\
c & 60.97 & 1.47 & 0.58 & 0.31  & 3.57 & 0.22
& 9.0  & 2.39  & 0.55
\end{tabular}
\end{ruledtabular}
\end{table}

Figure 4 shows our result for $^7$Be($d,n$)$^8$B cross section
using $V_{\rm pBe}$ of
Kim {\it et al}~\cite{Kim}., compared with the experimental data~\cite{Liu}.
The solid and dashed lines represent the calculated
results with $S_{\rm exp}=1.0$ and 0.849, respectively.
Parameter sets of the optical potentials at 4.08 and 4.26 MeV and the
single-particle potential of Kim {\it et al}. used in the calculation are
listed in Table 1 together.
At forward angles $\theta < 60^\circ$,
the calculated cross section well reproduces the data~\cite{Liu}
with the spectroscopic factor $S_{\rm exp}=0.849$, leading to
$S_{17}(0)= 21.36$ eV b.
It was found that deuteron breakup states play important roles not
only in determining distorting potentials but also in the transfer
process. In fact, when the deuteron breakup components
is set to zero in $\Psi_{\rm i}^{(+)}$, the resultant transfer cross
section is reduced by 10 \% at the forward angles and, more seriously,
the angular distribution cannot be reproduced correctly.
The components are obviously not included in the framework of the
standard DWBA.
We thus conclude that the three-body model approach is inevitable.

\begin{table}[ht]
\caption{Results of $S_{\rm exp}$, $b$, and $S_{17}(0)$ with different
$^8$B single particle models; see Eq.~(\protect\ref{S17}) for definition.}
\begin{ruledtabular}
\begin{tabular}{l|ccc}
 & $S_{\rm exp}$ & $b$ & $S_{17}(0)$ \\ \hline
Kim {\it et al.}~\protect\cite{Kim} & 0.849 & 0.809 & 21.36\\
Tombrello~\protect\cite{Tombrello} & 0.882 & 0.784 & 20.87\\
Robertson~\protect\cite{Robertson} & 0.864 & 0.794 & 20.93\\
Esbensen and Bertsch~\protect\cite{Esbensen} & 1.097 & 0.700 & 20.67
\end{tabular}
\end{ruledtabular}
\end{table}
We show in Table 2 the list of calculated $S_{\rm exp}$, $b$, and
$S_{17}(0)$ for different $V_{\rm pBe}$. One sees that the calculated
values of
$S_{17}(0)$ are almost consistent, which shows that, as mentioned
above, the present reaction is peripheral and the ANC method works well.
Taking account of the theoretical errors of ANC,
the adiabatic approximation (AD) and the finite-range correction (FRC)
and of the systematic error of the experimental data~\cite{Liu}
on the $^7$Be($d,n$)$^8$B cross section,
we obtain
$
S_{17}(0)= 20.96_{-0.3}^{+0.4}$ (ANC) $\pm0.63$ (AD) $\pm0.42$ (FRC)
$\pm2.7$ (expt) eV b,
in consistent with both the recommended values~\cite{Adelberger}
and the recent result $S_{17}(0)= 22.3\pm1.2$ eV b of
acccurate direct measurement~\cite{junghans}.

In summary, the present analyses based on the three-body model
are free from uncertainties of the optical potentials
in both the entrance and exit channels which were
the most essential problem in the previous DWBA analyses.
The deuteron breakup process in the incident channel is significant
for determining $S_{17}(0)$ within 5 \% error required from
the neutrino physics. We then conclude that
the three-body model approach is essential and necessary.
The present analyses provide a precise value
$S_{17}(0)= 20.96_{-1.3}^{+1.4}({\rm theor})\pm2.7({\rm expt})$ eV b;
the theoretical ambiguity of $S_{17}(0)$ is 6--7 \%
slightly beyond the required accuracy. However, we can reduce
the theoretical error to $\sim 2 \%$ coming from ANC only,
if we do the finite-range calculation
with accurate $\Psi_{\rm i}^{(+)}$ and $\Psi_{\rm f}^{(-)}$ derived by CDCC.
Such full-fledged calculations are highly expected.
In the present analyses, however, the experimental error (13 \%) is even
larger than the theoretical one.
It is expected that the peripheral properties, essential for the ANC method,
become insufficient, as the incident energy increases~\cite{Fernandes}.
Thus, accurate measurements on $^7$Be($d,n$)$^8$B and $^7$Li($d,d$)$^7$Li
at about a few tens of MeV and on $^7$Li($N,N$)$^7$Li at the half
the corresponding deuteron incident energy are highly expected;
the proton and deuteron elastic scattering are necessary to determine
the nucleon optical potential accurately.

The authors would like to thank T.~Motobayashi, T.~Kajino,
Y.~Watanabe, and M.~Kawai for helpful discussions.
This work has been supported in part by the Grants-in-Aid for
Scientific Research
of the Ministry of Education, Science, Sports, and Culture of Japan
(Grant Nos.~14540271 and 12047233).


\begin{thebibliography}{00}

\bibitem{Bahcall}
    J. N. Bahcall {\it et al.},
    Astrophys. J. {\bf 555}, 990 (2001) and references therein.

\bibitem{Adelberger}
    E. G. Adelberger {\it et al}.,
    Rev. Mod. Phys. {\bf 70}, 1265 (1998).

\bibitem{Xu}
    H. M. Xu {\it et al.},
    Phys. Rev. Lett. {\bf 73}, 2027 (1994).

\bibitem{Liu}
    Weiping Liu {\it et al.},
    Phys. Rev. Lett. {\bf 77}, 611 (1996).

\bibitem{Gagliardi}
    C. A. Gagliardi {\it et al.},
    Phys. Rev. Lett. {\bf 80}, 421 (1998).

\bibitem{Fernandes}
    J. C. Fernandes {\it et al.},
    Phys. Rev. C {\bf 59}, 2865 (1999).

\bibitem{Nunes1}
    F. M. Nunes {\it et al.},
    Nucl. Phys. {\bf A596}, 171 (1996).

\bibitem{Nunes2}
    F. M. Nunes {\it et al.},
    Nucl. Phys. {\bf A615}, 69 (1997); {\bf A627}, 747 (1997).

\bibitem{CDCC}
    M. Kamimura {\it et al.},
    Prog. Theor. Phys. Suppl. {\bf 89} (1986);
    N. Austern {\it et al.},
    Phys. Rep. {\bf 154}, 125 (1987).

\bibitem{Piya}
    R. A. D. Piyadasa {\it et al.},
    Phys. Rev. C {\bf 60}, 044611 (1999).

\bibitem{CDCC-foundation}
    N. Austern {\it et al.},
    Phys. Rev. Lett. {\bf 63}, 2649(1989);
    N. Austern {\it et al.},
    Phys. Rev. C {\bf 53}, 314 (1996).

\bibitem{Surrey}
    J. A. Tostevin {\it et al.},
    Phys. Rev. C {\bf 66}, 024607 (2002);
    A. M. Moro {\it et al.},
    Phys. Rev. C {\bf 66}, 024612 (2002) and references therein.

\bibitem{n+A.dat}
    H. D. Knox and R. O. Lane,
    Bulletin of the American Phys. Soc. {\bf 23}, 942 (1978).

\bibitem{d+A.dat}
  S. N. Abramovich {\it et al.},
  Izv. Rossiiskoi Akademii Nauk, Ser. Fiz. {\bf 40}, 842 (1974).

\bibitem{Johnson-Soper}
    R. C. Johnson and P. R. J. Soper,
    Phys. Rev. C {\bf 1}, 976 (1970).

\bibitem{fr-corr}
    J. A. Buttle and L. J. B. Goldfarb,
    Proc. Phys. Soc. {83}, 701 (1964).

\bibitem{Kim}
    K. H. Kim {\it et al.},
    Phys. Rev. C {\bf 35}, 363 (1987).

\bibitem{Tombrello}
    T. A. Tombrello,
    Nucl. Phys. {\bf 71}, 459 (1965).

\bibitem{Robertson}
    R. G. H. Robertson,
    Phys. Rev. C {\bf 7}, 543 (1973).

\bibitem{Esbensen}
    H. Esbensen and G. F. Bertsch,
    Nucl. Phys. {\bf A600}, 66 (1996).

\bibitem{n+A.dat2}
    H. D. Knox {\it et al.},
    Nuclear Science and Engineering {\bf 69}, 223 (1979).

\bibitem{junghans}
    A. R. Junghans {\it et al.},
    Phys. Rev. Lett. {\bf 88}, 041101 (2002).

\end{thebibliography}
\end{document}